\magnification =1200 


\hoffset=0pt
\voffset=0pt
\hsize=6.5 true in 
\vsize=8.5 true in  
\baselineskip=2.9ex


 at 6 true pt

\def\ds{\rat{d\s}{|\s|}\, }
\def\ST{{\sss,\t}}

\def\5#1{{\cal#1}}

\def\aaa{\alpha}

\def\ve{\varepsilon}
\def\fff{\phi}

\def\ppp{\pi}

\def\sss{\sigma} 

\def\ttt{\tau}

\def\yyy{\psi}

\def\Y{\Psi}

\def\({\left(}
\def\){\right)}
\def\[{\left[}
\def\]{\right]}
\def\lb{\left[}
\def\rb{\right]}

\def\0#1{(#1)}
\def\1#1{\hat#1}
\def\v1#1{{\hat#1}}
\def\2#1{\tilde#1}
\def\3#1{{\bf#1}}
\def\8{\infty}
\def\={\equiv}

\def\c#1{{\cal#1}}

\def \db#1{\raise.8ex\hbox{-}\kern-1.5ex d \kern .3ex#1}
\def \dbb#1#2{\raise.8ex\hbox{-}\kern-1.5ex d \kern .3ex^#1{\bf#2}\,\,}
\def\dd#1{_{\sst#1}}

\def\ea{\eqalign}

\def\inc#1{\int_C\!\!\dbb 3p\,\,}  
\def\incp#1{{\int_{{C\dd+}}\!\!\dbb 3p\,\,}}

\def\i1#1{\int_{-\infty}^\infty d#1\,\,} 
\def\ii{^{-1}}

\def\intt{\int\!\!\!\int}
\def\ir{\int_{-\infty}^\infty}

\def\la{\langle\, }

\def\lra{\leftrightarrow} 
 
\def\m#1{\,e^{-2\ppp\,#1}\,}

\def\notto{\ /\kern-2.2ex \to }
\def\0#1{(#1)}

\def\p#1{\,e^{2\ppp\,#1}\,}

\def\q{\quad} 
\def\qq{\qquad}

\def\ra{\,\rangle} 
\def\rat#1#2{{{#1}\over{#2}}}
   
\def\ref{\null}

\def\s#1{\sum_{#1\in\zz}}

\def\sst#1{{\scriptscriptstyle#1}}
\def\st#1{{\sl{#1}\/}} 
\def\ST{{s,t\,}}

\def\sv#1{\vglue#1ex}
  
\def\t#1{\tilde#1}

\def\v#1{{\bf#1}}

\def\x{\null}

\def\and{\q\hbox{and}\q}

\def\cl{\centerline}

\def\FT{Fourier transform} 
 
\def\ie{i.e., }
\def\iff{if and only if}
 
\def\ip{inner product}
\def\lhs{left--hand side}    
\def\n{\noindent}

\def\rhs{right--hand side}

\def\where{\q\hbox{where}\q}

\def\wrt{with respect to}

\newcount\eq\eq=1

\def\e#1{\eqno(#1.\the\eq)\global\advance\eq by 1$$}
\def\ee#1#2{\eqno(#1.#2.\the\eq)\global\advance\eq by 1$$}
\def\et{\eqno(10.\the\eq)\global\advance\eq by 1$$}
\def\en{\eqno(\the\eq)\global\advance\eq by 1$$}

\def\monthname{\ifcase\month\or January\or
   February\or March\or April\or May\or June\or July\or August
   \or September\or October\or November\or December\fi}

\def\RA{\kern-2pt\rightarrow\kern-2pt}
     
\def\har#1{\smash{\mathop{\hbox to .2in{\rightarrowfill}}
      \limits^{#1}}}
     
\def\harr#1{\smash{\mathop{\hbox to .5in{\rightarrowfill}}
      \limits^{#1}}}
      
\def\harl#1{\smash{\mathop{\hbox to .5in{\leftarrowfill}}
      \limits^{#1}}}

\def\frame#1#2{\cl{\vbox{\hrule height .1pt
                    \hbox{\vrule width .1pt\kern 10pt
                    \vbox{\kern 10pt
                    \vbox{\hsize #1cm\noindent#2}
                    \kern 10pt}
                    \kern 10pt\vrule width .1pt}
                    \hrule height 0pt depth.1pt}}}


\def\ST{{\sss,\ttt}}
\def\s{{\sigma}}
\def\t{{\tau}}

\def\c{{\chi}}
\def\y{{\psi}}

\def\5#1{{\cal#1}}


\cl{{{\bf Wavelet Filtering with the Mellin Transform$\,^*$}}
\vfootnote*{Supported by AFOSR grants  Nos.\  F49620-93-C-0063 
and F4960-95-1-0062.}  }
\sv1
\cl{(Appeared in \st{Applied Mathematics Letters} {\bf 9} \#5, 1996.)}

\sv2

\cl{\bf Gerald Kaiser}
\cl{Department of Mathematical Sciences}
\cl{University of Massachusetts-Lowell}
\cl{Lowell, MA 01854}\sv1

\sv1

\cl{December, 1995}

\sv4

\cl{\bf ABSTRACT}\sv1

{\narrower\narrower
\n It is shown that any convolution operator  in the time domain can be
represented \st{exactly} as a multiplication operator in the time-scale (wavelet)
domain.  The Mellin transform gives a one-to-one correspondence between
\st{frequency filters} (system functions)  and \st{scale filters}
(multiplication operators in the scale domain),  subject to the convergence of
the defining integrals. The usual wavelet reconstruction theorem is a special case.  
Applications to the denoising of random signals are proposed.  It is argued that the
present method is more suitable for removing the effects of atmospheric turbulence
than the conventional procedures because it is ideally suited for resolving spectral
power laws.

\sv2

\n {\bf Keywords:} Non-stationary processes: Time-scale representations.

\par}

\sv4

{\n \bf 1. Scale Filtering in the Wavelet Domain}

\n Time-scale analysis of signals is similar to their time-frequency analysis, but
with the frequency replaced by a scale parameter [1].  Instead of having to
choose a basic \st{window} in time, one must begin with a basic \st{wavelet}
$\yyy\0t$.  We define the \st{wavelet family} of $\yyy\0t$ as the
two-parameter family of functions
$$
\y_\ST\0t=\y(\s t-\t)\qq\hbox{with } \s, \t\hbox{ real and }\s\ne 0.
\en
Note that this differs from the usual convention
$\yyy_{s, \t'}\0t=|s|^{-1/2}\yyy((t-\t')/s)$. As will be seen below, \x(1) is
equivalent but formally simpler. Whereas the usual scale parameter $s$ can
be interpreted as \st{time scale}  ($|s|\gg 1$ means coarse scale and $0<|s|\ll 1$
means fine scale), our parameter $\s=1/s$ can be interpreted as \st{frequency
scale:} Large $\s$ isolates high frequencies, while small
$\s$ isolates low frequencies.  Thus we may \st{identify} $\s$ with the
frequency band passed by the wavelet $\yyy_\ST\0t$, and $\0\ST$ can be
regarded as a pair of time-frequency parameters, with the understanding
that now ``frequency'' means ``frequency scale.''  That is, instead of
\st{shifting} ($f\to f+f_0$), our frequencies now multiply ($\s\to\s\s_0$).
The absence of the factor $|s|^{-1/2}=|\s|^{1/2}$ will be seen to be unimportant
and, moreover, to lead to simpler reconstruction formulas.  Our parameter $\t$
also differs from the above $\t'$ by $\t=\s\t'$. That is, $\t$ is a time shift in the
\st{scaled} time.  (First scale, then shift rather than the other way around.) The
wavelet transform of a signal $\c\0t$ \wrt{} the family
$\yyy_\ST$ is defined as the \ip
$$
\2\c\0\ST\=\la\y_\ST\,,\c\ra=\i1t\y(\s t-\t)^*\c\0t.
\en
Inserting the Fourier expansion of $\y(\s t-\t)$ gives
$$\ea{
\2\c\0\ST&=\i1t\i1f \m{if(\s t-\t)}\1\y\0f^*\c\0t\cr
&=\i1f \p{if\t}\1\y\0f^*\1\c(f\s)\cr
&=\5F\ii\lb \1\y\0f^*\1\c(f\s)\rb\0\t,
\cr}\en
where $\1\y\0f$ denotes the \FT{} of $\y\0t$ and $\5F\ii$ is the
inverse \FT{} operator. Thus by Plancherel's theorem,
$$\ea{
\i1\t |\2\c\0\ST|^2&=\i1f |\1\yyy\0f|^2\, |\1\c(f\s)|^2\cr
&=\rat1{|\s|} \i1f |\1\yyy(f/\s)|^2\, |\1\c\0f|^2.
\cr}\en
Suppose we integrate both sides over  $\s$  with an arbitrary positive
\st{weighting function} $w\0\s$.  (For reasons to become clear below, we need
all scales $\s\ne 0$. For some purposes, $\s>0$ will do; this can be arranged
simply by choosing $w\0\s=0$ for $\s<0$.)  Thus
$$\ea{
\i1\s w\0\s\i1\t |\2\c\0\ST|^2
&=\ir\ds w\0\s\i1f |\1\yyy(f/\s)|^2\,|\1\c\0f|^2\cr
&=\i1f |\1\c\0f|^2\ir\rat{d\s}{|\s|} w\0\s\, |\1\yyy(f/\s)|^2,
\cr}\en
where the integral over $\s$ is understood to exclude $\s=0$ (integrate over
$0<\ve\le|\s|<\8$, then take the limit $\ve\to 0$). Let
$$
\Y\0f\= |\1\yyy\0f|^2
$$
denote the spectral density of $\y\0t$, and define
$$\ea{
W\0f&\=\ir\ds w\0\s\, \Y(f/\s)\=(w\bullet \Y)\0f\cr
&=\ir\ds \Y\0\s \,w(f/\s)\=(\Y\bullet w)\0f,
\cr}\en
where the second line is obtained from the first by the substitution $\s\to
f/\s$, assuming that $f\ne 0$. (The ``DC component'' $f=0$ gets special
treatment throughout wavelet theory, as does $\s=0$, and for similar reasons.)
Then (5) becomes
$$
\intt d\s\,d\t\, w\0\s \, |\2\c\0\ST|^2=\i1f W\0f\, |\1\c\0f|^2.
\en
The \lhs{} in \x(7) is a \st{scale-weighted norm} (``energy'')  of the signal,
whereas the \rhs{} is a \st{frequency-weighted norm} with weight function
$W\0f$.  (For example, \st{Sobolev norms}  are defined with $W\0f=(1+f^2)^s$
for some power $s$.) Then \x(7) is a kind of ``Plancherel theorem'' equating the
two weighted norms. Like the Plancherel theorem, it can be \st{polarized} to
give the analog of Parseval's relation for two signals $\fff\0t$ and $\c\0t$:
$$
\intt d\s\,d\t\ \2\fff\0\ST^*w\0\s \,  \2\c\0\ST
=\i1f \1\fff\0f^*W\0f\,  \1\c\0f.
\en
This shows that the \st{convolution operator} $\5W$ with system function (or
``symbol'') $W\0f$ can be expressed in the wavelet domain as
$$
(\5W\c)\0t\=\i1f \p{ift} W\0f\,\1\c\0f
=\intt d\s\,d\t\ \y_\ST\0t\, w\0\s \,  \2\c\0\ST.
\en
\st{Thus, to represent $\5W$ in the wavelet domain, we need merely to replace
$\1\c\0f$ with $\2\c\0\ST$, $W\0f$ with $w\0\s$, and the ``basis vectors''
$\p{ift}$ of the \FT{}  with the ``basis vectors'' $\y_\ST\0t$ of the wavelet
transform!} 

\sv1

In deriving \x(9) we assumed that $w\0\s$ (and therefore, by \x(6), also
$W\0f$) is nonnegative. This assumption was made for technical reasons, since
then all the integrands occurring in Equations \x(5)--(8) are nonnegative and
the integrals therefore either converge absolutely or diverge to infinity, making
it unnecessary to worry about conditional convergence. We now relax this
assuption and allow $w\0\s$ (hence also $W\0f$) to be  \st{complex-valued,} 
with the understanding that the integrals may no longer converge absolutely. 
Then \x(9) shows that operations normally performed in
the frequency domain can now also be performed in the time-scale
domain, \st{provided we can find a weighting function $w\0\s$ which gives
$W\0f$ by} \x(6). So the first problem that must be addressed is:  \st{Given
$W\0f$, solve} \x(6) \st{for $w\0\s$.}  We now show that this problem has a
unique solution, subject to the \st{admissibility condition} that its defining
integrals converge.

The key is to note that  Equation \x(6) defines $W\0f$ as a
\st{scaling convolution} of $w\0\s$ with $\Y\0f$, in the sense that the usual
difference variable $f-\s$ is  replaced by the quotient $f/\s$ and the
translation-invariant (Lebesgue) measure $d\s$ is replaced by the
scaling-invariant measure $d\s/|\s|$. Just as convolutions are converted into
products by the \FT, scaling convolutions are converted into products by the
\st{Mellin transform} [2]. The latter takes a function $F\0\s$ of a
\st{positive} variable $\s>0$ into a function $\breve F\0p$ of a \st{complex}
variable $p$ by
$$
\5M\{F\0\s\}\0p=\int_0^\8\rat{d\s}\s\ \s^{-p}\,F\0\s\=\breve F\0p,
\en
and the original function is reconstructed by integrating over a contour $C$
in the complex plane of the ``power parameter'' $p$ going from $c-i\8$ to
$c+i\8$, for an appropriate choice of $c$ which depends on $F\0\s$:
$$
F\0\s=\rat1{2\ppp i}\int_C dp\
\s^p\,\breve F\0p\=\5M\ii\{\breve F\0p\}\0\s.
\en
To solve \x(6) for $w\0\s$ using the Mellin transform, we must separate the
positive- and negative-frequency components of $W\0f$, since the Mellin
transform uses only $\s>0$.  Let us assume that
$$
\1\yyy\0f=0 \q\hbox{for }  f<0, \q\hbox{hence}\q \Y\0f=0 \q\hbox{for }  f<0.
\en
Then the wavelet $\y\0t$ is necessarily \st{complex.} It is an \st{analytic
signal} in the sense of Gabor [3], extending analytically to the upper half 
of the complex time plane. (See also  [1], Section 9.3.) Let $U\0\s$ be the unit
step function
$$
U\0\s=\cases{1 & if $\s>0$\cr 0 & if $\s<0$,\cr}
$$
and write
$$
w\0\s=U\0\s w\0\s+U(-\s)w\0\s=w\dd+\0\s+w\dd-(-\s),
\en
where
$$
w\dd\pm\0\s\=U\0\s w(\pm\s)=0\hbox { for } \s<0.
\en
By \x(12),  \x(6) becomes (for $f\ne 0$)
$$
W\0f=\int_0^\8\rat{d\s}\s\ \Y\0\s \,w(f/\s)=W\dd+\0f+W\dd-(-f),
\en
where
$$
W\dd\pm\0f=\int_0^\8\rat{d\s}\s\ \Y\0\s \,w\dd\pm( f/\s)
=(\Y\bullet w\dd\pm)\0f=0\hbox { for } f<0.
\en
Therefore the positive-frequency component $W\dd+\0f$ of $W\0f$ depends
only on the positive-scale component $w\dd+\0\s$ of $w\0\s$ and the
negative-frequency component $W\dd-(-f)$  depends only on the
negative-scale component $w\dd-(-\s)$.  This decoupling was the purpose of
the assumption \x(12) above.  Applying  the Mellin transform to the scaling
convolutions \x(16), we obtain
$$
\breve W\dd\pm\0p=\5M\ii\lb\Y\bullet w\dd\pm\rb\0p
=\breve \Y\0p\,\breve w\dd\pm\0p.
\en
This gives a formal expression for $w\dd\pm\0\s$ in terms of the transform of
$W\dd\pm\0f$:
\sv2
$$
w\dd\pm(\s)
=\rat1{2\ppp i}\int_C dp\ \s^{-p}\,\rat{\breve W\dd\pm\0p}{\breve\Y\0p},
\qq \s>0.
\en

\n We give three examples.  First, let $n$ be an integer and
$$
w\0\s=\s^n.
$$
Then
$$
w\dd\pm\0\s=U\0\s\, (\pm\s)^n,
\en
and \x(16) gives
$$
W\dd\pm\0f=U\0f \, (\pm f)^n\int_0^\8\rat{d\s}\s\ \Y\0\s\,\s^{-n}
=\,U\0f \, (\pm f)^n\,\breve\Y\0n.
\en
Hence
$$
W\0f=\breve\Y\0n\[U\0f\,f^n+U(-f)\,f^n\]=\breve\Y\0n \, f^n.
\en
That is, \st{pure integer powers are invariant under the correspondence
$w\0\s\lra W\0f$ between the frequency and scale domains, except for the
renormalization constant $\breve\Y\0n$.} The differential operator
$$
P\0D=\sum_{n=0}^N a_n D^n, \where D=\rat d{d t},
$$
is represented in the frequency domain as multiplication by
$$
W\0f=\sum_{n=0}^N a_n\, (2\ppp i f)^n.
\en
Hence, by \x(21),  it is represented in the wavelet domain as multiplication by
$$
w\0\s=\sum_{n=0}^N\rat{a_n (2\ppp i\s)^n}{\breve\Y\0n}.
\en
In order for this representation to be defined, we must have
$$
0<\breve\Y\0n\=\int_0^\8\rat{d\s}\s\  \Y\0\s\,\s^{-n}<\8\qq
\hbox{for all} n\hbox {with} a_n\ne 0.
\en
We say that the operator $P\0D$ is \st{admissible} if the condition \x(24) holds.
In particular, the identity operator $P\0D\=1$  is represented in frequency by
$W\0f\=1$, hence
$$
w\0\s=\rat1{\breve\Y\00}
$$
and \x(9) becomes
$$
\c\0t=\rat1{\breve\Y\00}\intt d\s d\t\,\y_\ST\0t\,\2\c\0\ST.
\en
This is the usual \st{wavelet reconstruction formula,} which inverts
the wavelet transform.  It shows that \st{the identity operator is  admissible
\iff}
$$
\breve\Y\00\=\int_0^\8\rat{d\s}\s\ |\1\y\0\s|^2<\8,
\en
which is the usual admissibility condition  [1] for the wavelet $\y\0t$.  Indeed, 
\x(18) indicates the following general \st{admissibility condition for convolution
operators:}  
\sv2

\cl{\frame{11}{In order for the  convolution operator with
system function  $W\0f$ to be representable in the wavelet domain of
the wavelet $\y\0t$, the Mellin transforms $\breve W\dd\pm\0p$ of
$W\dd\pm\0f$ must be analytic along an appropriate contour $C$ parallel to
the imaginary axis such that $\breve\Y\0p$ is also analytic with no 
zeros  along $C$ and the integral \x(18)  converges.}}

\sv2

\n This example suggests that wavelet methods may be used instead of Fourier
methods to solve differential equations.   To make this precise, one must
examine the admissibility condition of the given operator and the \st{range} of
the wavelet transform, \ie the set of all functions $\2\c\0\ST$ obtained from
$\c\0t$ as the latter is allowed to range over the  space of functions of interest.
(Not every function $F\0\ST$ of time and scale is the
wavelet transform of some time signal $\c\0t$; $F\0\ST$ must satisfy a
\st{consistency condition} related to the fact that the ``basis vectors''
$\y_\ST\0t$ are redundant [1].)

\sv1

Our second example is 
$$
w\0\s=|\s|^p,
\en
where $p$ is now any power (not necessarily an integer).
Then
$$
w\dd\pm\0\s=U\0\s\,\s^p,
\en
and \x(16) gives
$$
W\dd\pm\0f=U\0f \, f^p\int_0^\8\rat{d\s}\s\ \Y\0\s\,\s^{-p}
=U\0f \, f^p\,\breve\Y\0p.
\en
Hence
$$
W\0f=\breve\Y\0p\[U\0f\,f^p+U(-f)\,(-f)^p\]=\breve\Y\0p \, |f|^p.
\en
Like the pure integer powers $\s^n$ of $\s$, the pure powers $|\s|^p$ of $|\s|$
are invariant under the correspondence $w\0\s\lra W\0f$, again with the
renormalization factor $\breve\Y\0p$.  Note that this representation of the
filter $W\0f=|f|^p$ (with $p=-\aaa$)  is implicit in [1], Section 9.3.
\sv1

Our final example is
$$
h\0\s=\rat1{\breve\Y\00}\cases{-i & if $\s>0$\cr i & if $\s<0$.\cr}
\en
This gives
$$
H\0f=\cases{-i & if $f>0$\cr i & if $f<0$,\cr}
\en
which is the system function of the \st{Hilbert transform}
$$
(\5H\c)\0t=\rat1\ppp P\ir \rat{du}{u-t}\,\c\0t,
\en
where $P$ denotes the principal part of the integral.  Thus 
$$
(\5H\c)\0t=\rat i{\breve\Y\00}\[\intt_{\s<0} d\s d\t\,\y_\ST\0t\2\c\0\ST-
\intt_{\s>0} d\s d\t\,\y_\ST\0t\2\c\0\ST\].
\en
This representation of $\5H$ is implicit in Theorem 3.3 of [1].
\sv1

The combination of Mellin and Fourier transforms has long been used
by harmonic analylists in connection singular integral operators, in particular to
obtain estimates [4].  But although the concrete representation of convolution
operators derived here would seem to be fundamental in the application of
wavelet analysis to signal processing, I was unable to find its existence in the
literature.   A \st{discrete} version of this problem in the framework of
multiresolution analysis was the subject of a recent paper by Beylkin and
Torr\'esani [5].

\sv4

{\bf\n 2. Applications}
\sv1

\n The time-scale representation \x(9)  can be applied, in  principle, to any
problem involving convolution operators. Its usefulness
depends on the nature of the system function $W\0f$ for the problem at
hand.  The above examples show that systems whose functions obey power laws
will have simple representations.  In particular, the problem of filtering out
the effects of atmospheric turbulence (denoising)  seems a likely candidate,
since it involves asymptotic power laws. There is strong empirical
evidence that denoising in the scale domain is superior to denoising in the
frequency domain because it tends to smooth out the spectrum without
smearing the scales, so that the exponents in the power laws are clearly
resolved [6]. 
\sv1

The above time-scale analysis was developed for one dimension (time). It can
be extended to several dimensions, like space-time, and in that case it may be
applied to \st{physical wavelet representations.}  Physical wavelets, which
were defined in  [1], are localized acoustic or electromagnetic waves (\ie
solutions of the scalar wave equation or of Maxwell's equations) which share
many of the properties of the one-dimensional wavelets. They are related to one
another by geometrical operations such as space-time translations, scalings, and
Lorentz transformations (boosting to moving coordinate systems), and they can
be used as ``building blocks'' to form arbitrary acoustic or electromagnetic
waves. Such wavelets have recently found natural applications in sonar and
radar [7].   With them, the problem of denoising atmospheric turbulence can be
approached from \st{first principles,} since the physical wavelets already fully
describe the deterministic space-time behavior (dynamics) of the
\st{mean} process.  Progress on this will be reported elsewhere.

\sv3

\parindent =0pt

{\bf References}
\sv1

1.  G.~Kaiser,  \sl A Friendly Guide to Wavelets, 
\rm Birkh\"auser,  Boston, 1994.
\sv1

2.  R.~ Courant and D.~Hilbert,  \st{Methods of Mathematical Physics, } Volume 1,
Interscience, New York, 1953.
\sv1

3.  D.~Gabor,  Theory of communication, 
\sl J. IEE \rm {\bf 93}(1946) (III), 429--457.
\sv1

4. R.R.~Coifman and Y.~Meyer,  On commutators of singular integrals and
bilinear  singular integrals, \st{Transactions of the AMS} {\bf 212}, 315--331
(1975).
\sv1

5. G.~Beylkin and B.~Torr\'esani, \st{Implementation of operators via filter
banks,} preprint, University of Colorado,  October 13, 1995.
\sv1

6. L.~Hudgins, C.A.~Friehe, and M.E.~Mayer, Wavelet transforms and
atmospheric turbulence, \st{Physical Review Letters} {\bf 71}, 3279--3282
(1993).
\sv1

7. G.~Kaiser, Physical wavelets and radar, to appear in the \st{IEEE
Antennas  and Propagation Magazine,}  Febraury, 1996.
\sv1

\bye